\documentclass[%
preprint,
 amsmath,amssymb,
 aps,
doublecolumn
]{revtex4-1}

\usepackage{graphicx}
\usepackage{dcolumn}
\usepackage{bm}

\begin{document}


\title{Quality factor enhancement of Nano electromechanical systems by capacitive driving beyond the resonance}


\author{T. Barois}
\author{S. Perisanu}
\author{P. Poncharal}
\author{P. Vincent}
\author{S. T. Purcell}
\author{A. Ayari}%
 \email{anthony.ayari@univ-lyon1.fr}
\affiliation{%
Institut Lumi\`ere Mati\`ere, UMR5306 Universit\'e Lyon 1-CNRS, Universit\'e de Lyon 69622 Villeurbanne cedex, France.}%

\date{\today}

\begin{abstract}
Nano electromechanical systems are considered as ultra sensitive devices for mass and force detection. Capacitive actuation is widely used in these devices but is known to degrade the quality factor of the resonator due to DC electrostatic damping. We report the enhancement of the quality factor of SiC vibrating nanowires detected nano optomechanically and electrically by applying an AC capacitive driving at a frequency above both the resonance frequency and the electrical cut-off frequency. Self-oscillations are demonstrated for optimal conditions. We developed an analytical model of the phenomenon and showed that it can lead to an improvement of the force sensitivity.
\end{abstract}

\maketitle


\section{\label{sec:level1}INTRODUCTION}

Nano electromechanical systems (NEMSs)\cite{Craighead}, where electrons are coupled to mechanical motion, have generated a tremendous interest over the past years as they are promising candidates for testing macroscopic quantum physics\cite{teufel2011sideband,o2010quantum,chan2011laser}, non-linear dynamics at the nanoscale \cite{PhysRevB.81.165440,eichler_nonlinear_damping} and for their capability in sensors applications. These systems have shown record limits of detection reaching yoctogram mass resolution\cite{chaste2012} and a sensitivity of 12 zN/$\sqrt{Hz}$\cite{moser2013ultrasensitive}. In these electromechanical devices, the continuous improvement of the electrostatic transduction and the elimination of different sources of damping will ultimately be limited by the electrostatic damping\cite{jourdan2007tuning} as it is already the case for some single electron devices \cite{lassagne2009coupling,PhysRevLett.108.175502} or semiconducting nanowire cantilevers\cite{barois2012ohmic}. It is possible to reduce this damping by lowering the device resistance\cite{barois2012ohmic} or by applying a DC offset to compensate for residual voltages originating form work function differences and contact potentials\cite{siria2012electron}, but such solutions are not always convenient. It has been recently theoretically proposed\cite{eriksson2015nonresonant} that electrostatic coupling can also lead to a decrease of damping and to self-oscillations by simply applying an AC signal to a suspended electro mechanical graphene resonator and taking into account non-linear effects. The driving frequency is not related to the resonance frequency but just needs to be higher than the inverse of the charge relaxation time.

This method is appealing as (i) the operating frequency range is much less stringent than for systems coupled to a cavity\cite{metzger2004cavity} or a resonator\cite{rocard1971dynamique,brown2007passive,mahboob2012phonon}
(ii) it needs no additional elements than the already present coupling capacitor and NEMS resistance, (iii) it works for single and double clamped geometries, (iv) it gives a simple mean to tune the damping and to improve the signal to noise ratio for force sensing\cite{rugar1991mechanical} as already demonstrated for instance for magnetometers \cite{thompson2009parametrically,kumar2015ultra}. In this article, an experimental proof of this capacitively induced decrease of a NEMS mechanical damping is presented. We demonstrate that this effect can be large enough to induce self-oscillations in a mechanical resonator. The model of Eriksson et al.\cite{eriksson2015nonresonant} is then extend as this effect raises interesting question about the influence of the electrostatic actuation itself to perform an accurate measurement of the damping in NEMS and about the noise it induces for cooling and sensing applications.

\begin{figure}
\includegraphics[width=12cm]{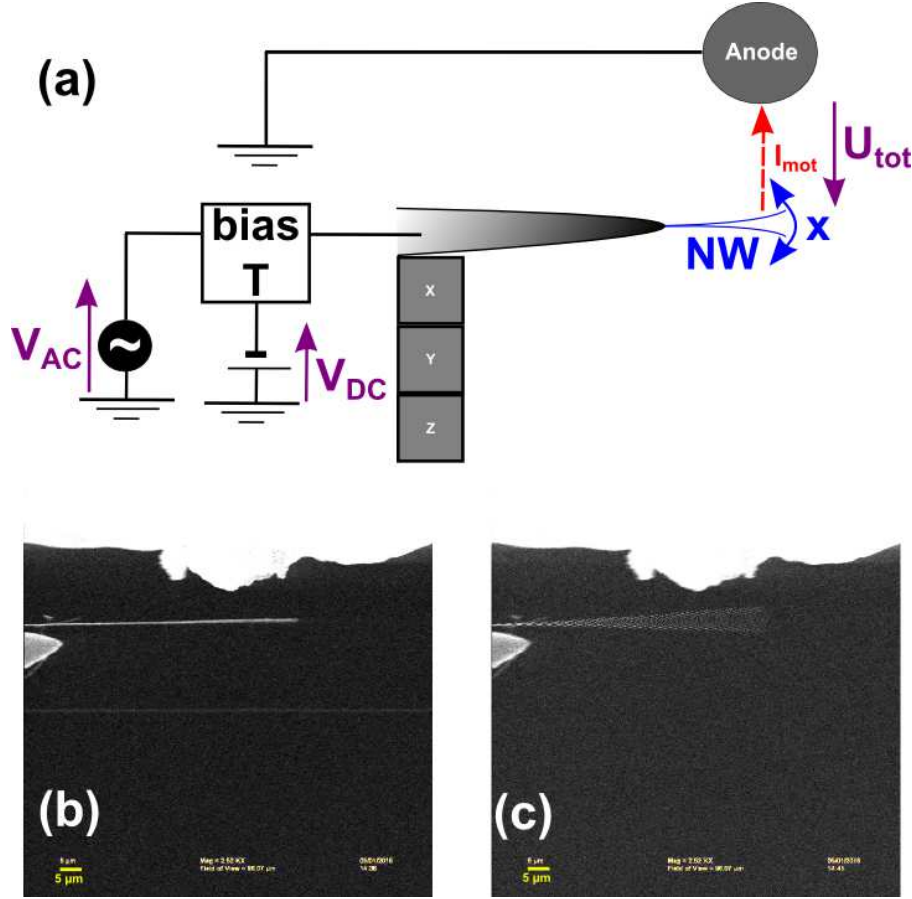}
\caption{\label{fig1} (a) Schematic of a vibrating SiC nanowire on a tungsten tip submitted to an AC voltage V$_{AC}$ or a DC voltage V$_{DC}$. (b) SEM image of the silicon carbide nanowire NW2 close to a gold counter electrode for V$_{AC}$ = V$_{DC}$ = 0 V. (c) SEM image of NW2 in self-oscillation for V$_{AC}$ = 19 V, V$_{DC}$ = 0 V and a driving frequency $\Omega/2\pi$ = 810 kHz well above the resonance frequency of 40 kHz.}
\end{figure}

\section{\label{sec:level2}RESULTS}
We studied 2 different SiC nanowires (samples NW1, length L = 34 $\mu$m, radius r = 75 nm; NW2, L = 86 $\mu$m, r = 90 nm) fixed at one end to W rigid tips and submitted to AC and DC capacitive forces (see Figure \ref{fig1}a). The measurements were performed in two different ultra high vacuum chambers one equipped with a scanning electron microscope (SEM) column and the other equipped with a nano-optomechanical set up. Piezo-inertial nanomanipulators are used to position the sample close to a counter electrode and in the focal point of the electron or light beam.

In the electron chamber the mechanical motion of the nanowire can be imaged as in standard electron microscopy or measured by recording the secondary electron signal in spot mode\cite{barois2012ohmic}. Figure \ref{fig1}b is a SEM image of NW2 close to the counter electrode. The resonance frequencies of NW2 were measured with a lock-in amplifier for an applied AC voltage of 50 mV. The first transverse mechanical eigenmode showed two polarizations at 39.83 kHz and 40.01 kHz where the degeneracy is lifted by either the clamping condition or the nanowire asymmetry. The second and third eigenmodes were at 230 kHz and 620 kHz. The width of the resonance peaks is about 5 Hz and outside these peaks no mechanical signal can be measured or observed with the SEM imaging. However when a higher AC signal above 18 V is applied, spontaneous mechanical vibrations where observed for driving frequencies ranging from 750 kHz to 850 kHz. The excitation of higher eigenmodes can be ruled out because the imaging of the mode doesn't show any nodes as observed in figure \ref{fig1}c. Non-linear high order parametric excitation of the first mode and parametric coupling between the first mode and high order mechanical mode\cite{mahboob2012phonon} seem dubious because of the discrepancy between the theoretical expected frequency and the observed frequency range for spontaneous oscillations. For this type of sample, the resistance is of the order of 1 G$\Omega$ and the capacitance in the fF range\cite{barois2012ohmic,barois2013role}. This corresponds to an inverse of the charge relaxation time of the order of  hundreds of kHz. So our observations are in agreement with the model of Eriksson et al.\cite{eriksson2015nonresonant} predicting self-oscillations for a driving frequency higher than the charge relaxation time.
\begin{figure}
\includegraphics[width=12cm]{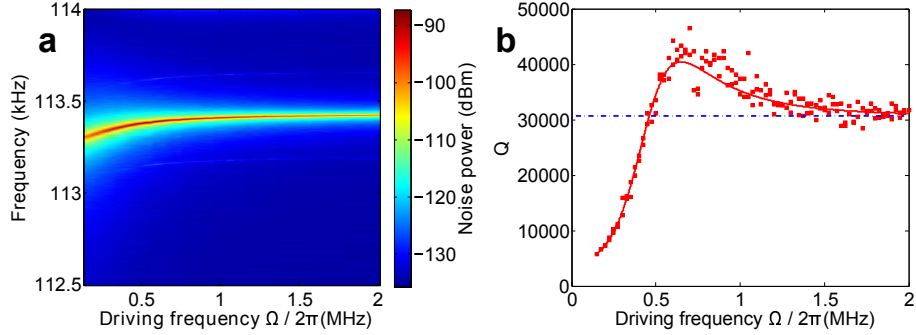}
\caption{\label{fig2} (a) Evolution of the displacement noise spectrum of NW1 as a function of the driving frequency $\Omega/2\pi$ for V$_{AC}$ = 10 V and V$_{DC}$ = 0 V. (b) Evolution of the quality factor Q as a function of the driving frequency. The solid line is a fit of the quality factor with equation \ref{gamafit}. The dashed line is the quality factor without an AC voltage.}
\end{figure}

We performed additional measurements with NW1 in our UHV optical set up in order to get a more quantitative agreement with theory. In the optical chamber, a 35 $\mu$W and 633 nm laser beam is strongly focused on the free end of the nanowire and the transmitted light is collected using a differential measurement based on a dual quadrant photodiode\cite{PhysRevLett.104.147203,gloppe2014bidimensional} giving access to the thermo-mechanical motion of the resonator with high signal to noise ratio. So, in this experimental set up, no AC probe signal is required to measure the mechanical motion and electron beam induced self-sustained oscillations\cite{vincent2007driving} will not interfere with our measurements. The first transverse mechanical eigenmode showed two polarizations at 113.434 kHz and 114.071 kHz. In the rest of the text, we will focus only on the low frequency polarization as the high frequency polarization shows the same behavior. An AC signal of 10 V is applied to the nanowire and swept back and forth between 150 kHz and 2 MHz in 25 kHz increments while recording the power spectrum density of the thermo-mechanical motion as plotted in figure \ref{fig2}a. A shift in frequency and an increase in vibration amplitude is observed when the frequency increases. The resonance peaks have been fitted with a lorentzian (see figure 1 in supplemental material). Figure \ref{fig2}b shows the variation of the quality factor defined as $\omega_0/\Gamma_{AC}$ where $\omega_0/2\pi$ is the resonance frequency and $\Gamma_{AC}$ the resonance width. The dashed line indicates the initial quality factor Q$_0$ = 30 600 in the absence of AC or DC voltages. At low driving frequencies, the quality factor is strongly reduced but increases with the frequency until it reach reach a maximum above 40 000 around 650 kHz. This maximum is significantly higher than the initial quality factor. At higher frequencies, Q tends to Q$_0$.

At first sight, this effect seems rather counter intuitive as the electrostatic interaction usually induces an increase in damping when a DC voltage is applied and AC actuation is very often used as a reliable mean to drive NEMS. Furthermore the possibility to increase or decrease the damping is reminiscent of coupling with a blue-detuned or red-detuned cavity whereas no cavity is involved in this system. The mechanical system can be modelized as a mechanical resonator capacitively connected to its environment.
\begin{equation}
\label{meca}
m\ddot{x}+m\Gamma_0\dot{x}+m\omega_0^2x = \frac{C'}{2}(U_{tot})^2+F_T(t)
\end{equation}
where x is the position in the transverse direction of the free end of the nanowire, m is the effective mass of the mechanical resonator, the dot refers to the derivative versus time t, $\Gamma_0$ is the intrinsic mechanical damping, $F_T(t)$ is the Langevin force coming from the thermal bath, $C'$ is the space derivative of the capacitance C formed between the resonator and the electrostatic environment, $U_{tot}$ is the total voltage difference between the point x and the ground (see Fig. \ref{fig1}a). This voltage is provided by a transmission line that can be roughly described by the following equation :
\begin{equation}
RC\dot{U}_{tot}+U_{tot} = -RC'U_{tot}\dot{x}-RC'\dot{U}_{tot}x+V_{AC}\cos\Omega t
\end{equation}
\begin{figure}
\includegraphics[width=12cm]{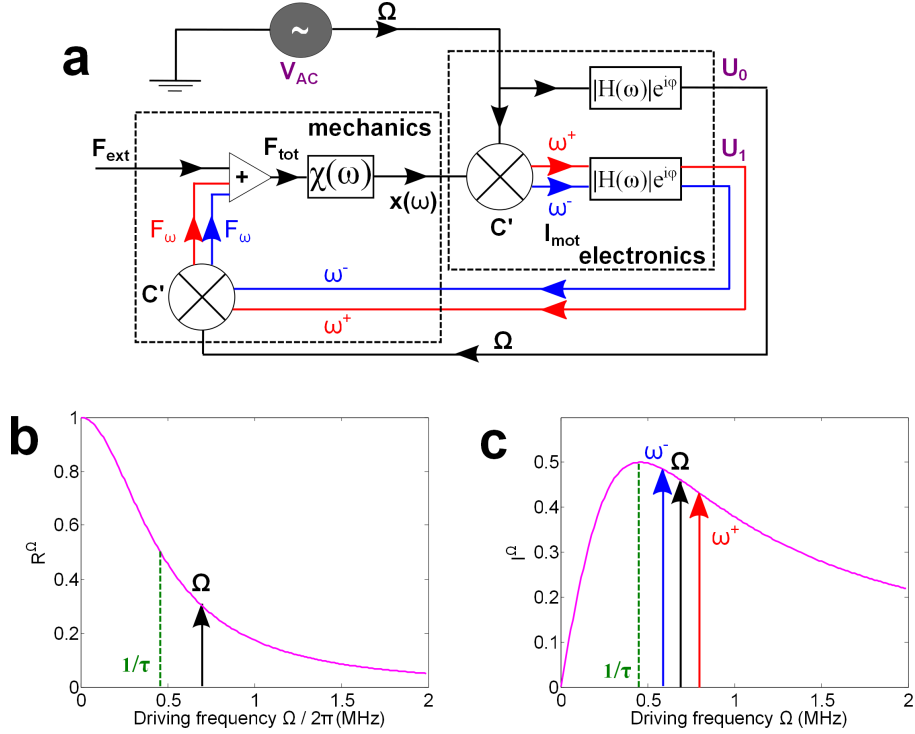}
\caption{\label{fig3} (a) Schematic of the mechanism involved in the quality factor enhancement induced by an AC voltage. F$_{ext}$ is an external force like the thermal noise force. F$_{tot}$ is the sum of the external force and the two feedback forces. $\chi(\omega)$ is the mechanical susceptibility of the nanowire. $H(\omega)e^{i\varphi}$ is the transfer function of the transmission line. U$_0$ and U$_1$ are calculated in the supplemental material. The two crossed circles represent signal mixers responsible for the non-linear effects and originating from the C' terms. (b) Evolution of the term $\Re^\Omega$ as un function of the driving frequency with the experimental parameters of NW1. The smaller the black arrow, the smaller the effect on the damping. (c) Evolution of the term $\Im^\Omega$ as un function of the driving frequency with the experimental parameters of NW1. The damping is reduced as long as the blue arrow is longer than the red arrow.}
\end{figure}
where R is the resistance of the resonator including the contact resistance, $V_{AC}$ is the applied AC voltage at frequency $\Omega/2\pi$. The important terms are those related to the change in capacitance when the nanowire moves and changes its distance from the counter electrode. This motion induces a motional current $C'U_{tot}\dot{x}+C'\dot{U}_{tot}x$. These non-linear terms mix the driving voltage at angular frequency $\Omega$ with the nanowire motion at angular frequency $\omega$. A redshifted (Stokes) electromagnetic wave at $\omega_- = \Omega-\omega$ and a blueshifted (anti-Stokes) electromagnetic wave at $\omega_+ = \Omega+\omega$ are then produced and down-mixed by the squared term of the electrostatic actuation in Eq. \ref{meca} (see Figure \ref{fig3} and supplemental material). This induces a feedback mechanism that modify the damping and resonance frequency of the nanowire. At low driving frequencies, the electrical circuit should behave as if a DC voltage was applied and so an increase in damping is expected. The electrical actuation is a low pass filter, so for frequencies beyond the inverse of the charge relaxation time  $\tau=RC$ (i.e. the cut-off frequency), the effect of the applied voltage on the apex of the nanowire should vanish and the initial damping should be recovered . However, in the intermediate regime of frequency, the actuation is dephased compared to the DC actuation and the amplitude of the electromagnetic actuation although smaller can still influence the resonator. This dephasing allows a decrease of damping for a wide frequency range. The effective mechanical equation is then :
\begin{equation}
\ddot{x}+\Gamma_{AC}\dot{x}+\omega_{AC}^2x = 0
\end{equation}
with :
\begin{equation}
\label{gamafit}
  \Gamma_{AC} = \Gamma_0 + g_1^2V_{AC}^2\Re^\Omega\frac{\Im^{\omega_+}-\Im^{\omega_-}}{2\omega} ; \Re^\Omega = \frac{1}{1+(\Omega\tau)^2}; \Im^\Omega = \frac{\Omega\tau}{1+(\Omega\tau)^2};  g_1 = \frac{C'}{\sqrt{2mC}}
\end{equation}
where $g_1$ is the electrostatic coupling strength for an applied voltage of 1V. It characterizes the strength of the electrostatic coupling. The blue-shifted term $\Im^{\omega_+}$ always leads to an increase in damping whereas the red-shifted term $\Im^{\omega_-}$ reduces the damping as long as $\omega<\Omega$. So this mechanism can not lead to a reduction in damping for a driving lower than the resonance frequency. For $\omega<<\Omega$ the damping can be approximated by :
\begin{equation}
\label{gamafit2}
  \Gamma_{AC} \approx \Gamma_0 + g_1^2V_{AC}^2\tau\frac{1-(\tau\Omega)^2}{(1+(\tau\Omega)^2)^3}
\end{equation}
For $\Omega\tau > 1$, the AC driving starts to decrease the damping. The quality factor enhancement is maximum for $\Omega=\sqrt{2}/\tau$ where the damping decreases by  $g_1^2V_{AC}^2\tau/27$. The data in Fig. \ref{fig2}b were fitted with Eq. \ref{gamafit2}. A very good agreement is observed between theory and experiment. It gives a charge relaxation time of 347 ns corresponding to a cut of frequency of 460 kHz, a maximum quality factor enhancement at 650.5 kHz and a coupling strength of 2098.5 rad/s/V corresponding to 333.98 Hz/V. Thus our system is in the weak coupling regime $g_1V_{AC}\tau \approx$ 0.007 $<<$1 and this explains why a rather important voltage of 10 V was necessary to observe this effect. A much more spectacular effect should be expected in the strong coupling regime. As a comparison, we performed a measurement of the resonance width as a function of the DC voltage without an AC voltage. As expected\cite{barois2012ohmic}, a square dependence in voltage was observed (see supplemental material) and fitted. From this fit and using a value of $\tau$ = 347 ns as determined previously, the coupling strength was estimated at 1622.5 rad/s/V in agreement with the g$_1$ value obtained from the AC experiment.

\begin{figure}
\includegraphics[width=12cm]{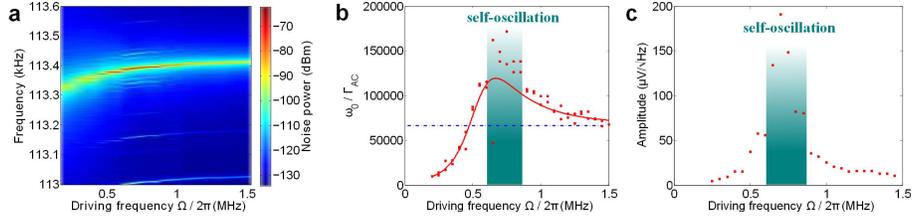}
\caption{\label{fig4} (a) Evolution of the displacement noise spectrum of NW1 after annealing  as a function of the driving frequency $\Omega/2\pi$ for V$_{AC}$ = 10 V and V$_{DC}$ = 0 V. Satellite peaks related to self-oscillation appear between 650 kHz and 850 kHz. (b) Evolution of $\omega_0/\Gamma_{AC}$ (which correspond to the quality factor Q outside the self-oscillation region) as a function of the driving frequency after annealing. The solid line is a fit of the quality factor with Eq. \ref{gamafit}. The dashed line is the quality factor without an AC voltage. (c) Evolution of the vibration amplitude at resonance as a function of the driving frequency.}
\end{figure}

An annealing of the nanowire was performed in order to remove surface adsorbates and increase the initial quality factor up to 66 700. An AC signal of 10 V was applied to the nanowire and swept back and forth between 200 kHz and 1.5 MHz in 50 kHz increments while recording the power spectrum density of the thermo-mechanical motion as plotted in Fig. \ref{fig4}a. Figure \ref{fig4}b shows a quality factor above 150 000 between 650 kHz and 850 kHz. The peak width could not be measured accurately in this frequency range due to the 1 Hz limitation of the spectrum analyzer bandwidth. If these points are excluded, a fit is still possible and gives similar values for g$_1$ and $\tau$ as before the annealing. The clear departure of the quality factor from the fit between 650 kHz and 850 kHz indicates that the nanowire self-oscillates. It can be pointed out here, that this is a true self-oscillation where the phase is free, contrary to the instability of the parametric oscillator where the phase is fixed by the AC generator.

\section{\label{sec:level3}DISCUSSION}
An interesting point, concerning our derivation of equation \ref{gamafit} is that it gives a simple expression of the damping valid for any $\Omega$ (except for $\Omega$=0 where a slight correction is needed). In particular, it appears from Eq. \ref{gamafit} that when $\Omega$ is close to the resonance frequency, the damping obtained with a capacitive AC driving is overestimated by $g_1^2V_{AC}^2\tau/(1+(\tau\omega_{AC})^2)^3$. This overestimation can be even higher when mixing technics are used. So, in the strong coupling regime and in the study of non-linear effects like the Duffing regime, the change in damping might have to be taken into account. Then, it can be expected that a change in quality factor with temperature could be simply related to for instance a change in the contact resistance. This could have implication also for the study of NEMS frequency noise by applying a capacitive driving close to the resonance and measuring the vibration noise spectrum as proposed by Zhang et al.\cite{zhang2014interplay}.

In our particular experiment at room temperature, although the transduction was improved, this doesn't lead to an improvement in the force sensitivity as the signal and thermal vibrations are amplified by an equal amount. Nevertheless at lower temperature or in systems where thermal vibrations are not the dominant source of noise a gain in sensitivity could be expected as long as the Johnson noise from the resistance is not too high. In the supplemental material, this additional force noise is estimated to be of the same order as the thermal force noise for NW1. The effective temperature was also calculated and we showed that, contrary to the DC electrostatic damping, the AC mechanism leads to heating of the mechanical mode. This heating can be in our case as high as 4 times the room temperature.

\section{\label{sec:level4}CONCLUSION}

In conclusion, we have shown experimentally that applying a capacitive AC signal can improve the quality factor of a nanowire mechanical resonator and lead to self-oscillations. The enhancement of the quality factor was maximum for a driving angular frequency at $\sqrt{2}$ the inverse of the charge relaxation time. We derived an analytical expression for this effect valid on a wide frequency range and calculated the influence of the AC driving on the force sensitivity of the device. This enhancement is not limited to single clamped geometry or nanowires but is more suitable for resistive sample such as semiconducting 2D sample like MoS$_2$ or samples contacted by a tunneling resistance. The simplicity of this mechanism opens up a new path for the simultaneous driving and ultimately synchronization\cite{barois2014frequency} of NEMS array as a single AC driving can lead to self-oscillations of all the devices even when they are not perfectly identical. This technique can be seen also as a mean to measure the charge relaxation time of a NEMS in the case where cross talk make difficult a direct electrical detection.

\begin{acknowledgments}
The authors thank the Plateforme Nanofils et Nanotubes Lyonnaise of the University Lyon1 and are especially grateful to the NOF team for their support. This work was supported by the  French National Research Agency (ANR) through its Blanc Programme (FOCUS, ANR-13-BDS10-0012-01).
\end{acknowledgments}

\nocite{*}

%

\end{document}